\documentclass[12pt]{iopart}
\usepackage{epsfig}
\usepackage{citesort}

\newcommand{\gtrsim}{\,\rlap{\lower3.7pt\hbox{$\mathchar\sim$}}
\raise1pt\hbox{$>$}\,}
\newcommand{\lesssim}{\,\rlap{\lower3.7pt\hbox{$\mathchar\sim$}}
\raise1pt\hbox{$<$}\,}

\begin{document}

\title{Decay of heavy Majorana neutrinos using the full
Boltzmann equation including its implications for leptogenesis}

\author{Anders Basb{\o}ll$^1$, Steen Hannestad$^{1,2}$}
\address{$^1$ Department of Physics and Astronomy, University of Aarhus,
Ny Munkegade, DK-8000 Aarhus C}
\address{$^2$Max-Planck-Institut f\"ur Physik
(Werner-Heisenberg-Institut), F\"ohringer Ring 6, D-80805
M\"unchen, Germany}

\ead{\mailto{andersb@phys.au.dk}, \mailto{sth@phys.au.dk}}
\date{{\today}}

\begin{abstract}
We have studied the two-body production and decay of a heavy,
right-handed neutrino to two light states using the full Boltzmann
equation instead of the usual integrated Boltzmann equation which
assumes kinetic equilibrium of all species. Decays and inverse
decays are inefficient for thermalising the distribution function
of the heavy neutrino and in some parameter ranges there can be
very large deviations from kinetic equilibrium. This leads to
substantial numerical differences between the two approaches.
Furthermore we study the impact of this difference on the lepton
asymmetry production during leptogenesis and find that in the
strong washout regime the final asymmetry is changed by 15-30\%
when the full Boltzmann equation is used.
\end{abstract}
%\pacs{14.60.Pq,95.35.+d,98.80.-k}
\maketitle

%%%%%%%%%%%%%%%%%%%%%%%%%%%%%%%%%%%%%%%%%%%%%%%%%%%%%%%%%%%%%%%%%%%%%%
\section{Introduction} %%%%%%%%%%%%%%%%%%%%%%%%%%%%%%%%%%%%%%%%%%%%%%%
%%%%%%%%%%%%%%%%%%%%%%%%%%%%%%%%%%%%%%%%%%%%%%%%%%%%%%%%%%%%%%%%%%%%%%

Leptogenesis is perhaps the most attractive model for generating
the matter-antimatter asymmetry in our Universe
\cite{Fukugita:1986hr} after inflation. The process generates a
lepton asymmetry via the production and subsequent decay of a
heavy Majorana neutrino. This lepton asymmetry is partially
converted to a baryon asymmetry via sphaleron processes
\cite{Kuzmin:1985} which break both $B$ and $L$, but conserve
$B-L$.

The fraction of $B-L$ that ends up in $B$ by the sphaleron
processes is given by $a_{sph}=28/79$ giving
$n_B=a_{sph}n_{B-L}=-a_{sph}n_L$ where the $n_{B-L}$ is created by
leptogenesis which gives a non zero value of $L$ (see for instance
\cite{Buchmuller:2004nz}).

In its simplest form (within the context of the see-saw model)
leptogenesis consists of adding three heavy right-handed neutrinos
to the standard model. In the hierarchical limit one of these
right handed neutrinos, $\nu_{R_1}$, is much lighter than the
other two and the leptogenesis mechanism consists essentially of
the process
\begin{equation}
\nu_{R_1} \to \cases{\nu_L + \phi \cr \overline{\nu_L} + \phi},
\end{equation}
where $\nu_L$ is a light, left-handed neutrino and $\phi$ the
Higgs. All light flavours behave identically, so you can think of
the other light flavours as included as a factor of $3$ in the
decay rate. Because of loop corrections there can be $CP$
violation in the decay, usually quantified by the asymmetry
parameter $\epsilon$. In the following we denote the heavy
neutrino by $R$, the light by $L$, and the Higgs by $H$.

This model has been studied extensively in the literature,
including deviations from the hierarchical limit, thermal
corrections etc \cite{Buchmuller:2004nz,Luty:1992un,Buchmuller:2000as,Hambye:2003rt,Giudice:2003jh,%
Buchmuller:2003gz,Giudice:1999fb,Hirsch:2006ft,Blanchet:2006dq,Blanchet:2006be,%
Abada:2006ea,Ota:2006xr,Nardi:2006fx,Abada:2006fw,Hambye:2005tk}.
However, all studies have used the integrated Boltzmann equation
to follow the evolution of the heavy neutrino number density and
the lepton asymmetry. This approach assumes Maxwell-Boltzmann
statistics for all particles as well as kinetic equilibrium for
the heavy species. This assumption is normally justified in
freeze-out calculations where elastic scattering is assumed to be
much faster than inelastic reactions. However, in the present
context, kinetic equilibrium in the heavy species would have to be
maintained by the decays and inverse decays alone. Therefore it is
not obvious that the integrated Boltzmann equation is always a
good approximation. Furthermore $1 \leftrightarrow 2$ processes
are generally inefficient for thermalization compared with $2
\leftrightarrow 2$ processes. For $2 \leftrightarrow 2$ deviations
from kinetic equilibrium are always of order 20\% or less
\cite{Hannestad:1999fj}, but for $1 \leftrightarrow 2$ processes
they can be very large (see for instance
Refs.~\cite{Madsen:1992am,Starkman:1993ik,Kaiser:1993bt,Hannestad:1997ai}
for a case where deviation from equilibrium is extremely large).

In this paper we investigate how the use of the full Boltzmann
equation affects the final lepton asymmetry in a simplified model
with only decays and inverse decays and resonant scattering. We
will return to the point of resonant scattering in due course. We
find that when $T \sim m_R$ the difference can be very large.
However, at small temperature where the inverse decay dominates
the difference decreases in magnitude to about 20\%.

%%%%%%%%%%%%%%%%%%%%%%%%%%%%%%%%%%%%%%%%%%%%%%%%%%%%%%%%%%%%%%%%%%%%%%
\section{The Boltzmann equation} %%%%%%%%%%%%%%%%%%%%%%%%%%%%%%%%%%%%%
%%%%%%%%%%%%%%%%%%%%%%%%%%%%%%%%%%%%%%%%%%%%%%%%%%%%%%%%%%%%%%%%%%%%%%

Here we study only the two-body decay of a heavy right-handed
neutrino to a light neutrino plus a Higgs. We do not include
thermal corrections to the particle masses, so that for instance
the process $H \to LR$ is not kinematically possible. We assume
that the asymmetry, represented by
\begin{equation}
\epsilon = - \frac{\Gamma - \bar\Gamma}{\Gamma + \bar\Gamma}
\end{equation}
is small, so that when we calculate anything with $R$, we can
assume identical distributions of $L$ and $\bar{L}$. We consider
only initial zero abundance of $R$.
  We use only single particle distribution
functions, in which case the Boltzmann equation for the heavy
species can be written as \cite{Kolb:1979qa,Luty:1992un}

\begin{eqnarray}
\frac{\partial f_R}{\partial t} - p H \frac{\partial f_R}{\partial
p} & = & \frac{1}{2E_R} \int \frac{d^3 p_L}{2E_L (2
\pi)^3}\frac{d^3 p_H}{2E_H (2 \pi)^3} (2 \pi)^4
\delta^4(p_R-p_L-p_H) \nonumber\\
&& \,\, \times \left[f_H f_L (1-f_R) (|M_{HL \to
R}|^2+|M_{H\bar{L} \to R}|^2)\right. \nonumber \\
&& \,\,\, \left. - f_R (1-f_L)(1-f_H) (|M_{R \to HL}|^2 +|M_{R \to
H\bar{L}}|^2) \right],
\end{eqnarray}
and for the light neutrino it is
\begin{eqnarray}
\frac{\partial f_L}{\partial t} - p H \frac{\partial f_L}{\partial
p} & = & \frac{1}{2E_L} \int \frac{d^3 p_R}{2E_R (2
\pi)^3}\frac{d^3 p_H}{2E_H (2 \pi)^3} (2 \pi)^4
\delta^4(p_R-p_L-p_H) \nonumber\\
&& \,\, \times \left[- f_H f_L (1-f_R) |M_{HL \to
R}|^2 \right. \nonumber \\
&& \,\,\, \left. + f_R (1-f_L)(1-f_H) |M_{R \to HL}|^2 \right],
\end{eqnarray}
with a similar equation for $\bar{L}$. The interesting Boltzmann
equations for the present purpose are those for $R$ and for $B-L$.
$H$ and $L,\bar{L}$ have gauge interactions which are very fast.
This means that $H$ can be described by a distribution in chemical
equilibrium, and that $L,\bar{L}$ can be described as
distributions in kinetic equilibrium,
\begin{eqnarray}
f_R & = &(1+e^{p_R/T})^{-1} \\
f_L & = &(1+e^{(p_L-\mu)/T})^{-1} \\
f_{\bar{L}} & = &(1+e^{(p_{\bar{L}}+\mu)/T})^{-1},
\end{eqnarray}
with $\mu/T = 3 (n_L - n_{\bar{L}})/T^3 + {\cal O}((\mu/T)^3)$.

Using CPT-invariance, following the idea of \cite{Kolb:1979qa}, we
find
\begin{eqnarray}
 |M_{R \to HL}|^2& = & |M_{H\bar{L} \to R}|^2=1-\epsilon \nonumber\\
|M_{R \to H\bar{L}}|^2 &=& |M_{HL \to R}|^2=1+\epsilon
\end{eqnarray}
This simplifies the Boltzmann equations.

For the heavy neutrino the Boltzmann equation can be simplified to
\cite{Starkman:1993ik}
\begin{eqnarray}
\frac{\partial f_R}{\partial t} & - & p H \frac{\partial
f_R}{\partial p} \\ && = \frac{m_R \Gamma_{\rm tot}}{E_R p_R}
\int_{(E_R-p_R)/2}^{(E_R+p_R)/2} dp_H \left[f_H f_L (1-f_R) - f_R
(1-f_L)(1-f_H) \right], \nonumber \label{eq:R}
\end{eqnarray}
where $\Gamma_{\rm tot} = \Gamma_{R \to LH}+ \Gamma_{R \to \bar{L}
H} = \Gamma + \bar{\Gamma}$ is the total rest frame decay rate. The
corresponding equation for $L-\bar{L}$ \footnote{Note that we ignore
the sphaleron $L$ to $B$ conversion during the decay process because
we track only $L-\bar{L}$. Including it would have only a modest
effect on our numerical results.} is
\begin{eqnarray}
\frac{\partial (f_L - f_{\bar{L}})}{\partial t} & - & p H
\frac{\partial (f_L-f_{\bar{L}})}{\partial p} \\ && = -\frac{m_R
\Gamma}{2 E_L p_L} \int_{\frac{m_R^2}{4p_L}+p_L}^{\infty} dE_R
\left[(f_H + f_R)F^- \right. \nonumber \\
&& \,+ \left. \epsilon\bigg(-2(1+f_H)f_R + (f_R-f_H(1-2f_R))F^+
\bigg )\right], \label{eq:L}
\end{eqnarray}
to first order in $\epsilon$ and with
\begin{eqnarray}
F^+ & = & f_L + f_{\bar{L}} = \frac{2}{1+e^{p/T}} + {\cal
O}((\mu/T)^2) \\
F^- & = & f_L - f_{\bar{L}} = \frac{2 e^{p/T}}{(1+e^{p/T})^2}
\frac{\mu}{T} + {\cal
O}((\mu/T)^3) \\
\end{eqnarray}

Since we have so far only included $2 \leftrightarrow 1$
processes, Eq.\ \ref{eq:L} suffers from the well-known problem of
lepton asymmetry generation even in equilibrium
\cite{Kolb:1979qa}. To remedy this problem the resonant part of
the $L H \leftrightarrow \bar{L}H$ must be included. To lowest
order in $\epsilon$ this amounts to adding the term $2 \epsilon (1
- f_R)f_H F^+$ in Eq.\ \ref{eq:L} \cite{Giudice:2003jh}, so that
the final form of the equation for the lepton asymmetry is
\begin{eqnarray}
\frac{\partial (f_L - f_{\bar{L}})}{\partial t} & - & p H
\frac{\partial (f_L-f_{\bar{L}})}{\partial p} \\ && = -\frac{m_R
\Gamma}{2 E_L p_L} \int_{\frac{m_R^2}{4p_L}+p_L}^{\infty} dE_R
\left[(f_H + f_R)(F^-+\epsilon F^+)-2 \epsilon f_R (1+f_H)\right],
\nonumber \label{eq:L2}
\end{eqnarray}

This equation does not exhibit any lepton asymmetry generating
behaviour in thermal equilibrium because $\left[(f_H +
f_R)(F^-+\epsilon F^+)-2 \epsilon f_R (1+f_H)\right] = 0$
explicitly.

To first order in $\mu/T$ equations Eq.\ \ref{eq:R} and
\ref{eq:L2} can be easily integrated numerically for given values
of $m_R$ and $\Gamma$. If Maxwell-Boltzmann statistics and kinetic
equilibrium for $R$ are assumed the equations can be further
simplified. The integrated Boltzmann equations are then

\begin{eqnarray}
\dot{n}_R + 3 H n_R & = & - \langle \Gamma \rangle (n_R -
n_{R,{\rm eq}}) \\
\dot{n}_{L-\bar{L}} + 3 H n_{L-\bar{L}} & = & \epsilon \langle
\Gamma \rangle (n_R - n_{R,{\rm eq}}) + n_{L-\bar{L}}
\frac{\Gamma}{4} {\cal K}_1(m_R/T) \frac{m_R^2}{T^2},
\end{eqnarray}
where $z \equiv\frac{m_R}{T}$ and $\langle \Gamma \rangle =
(\Gamma + \bar{\Gamma}) {\cal K}_1(z)/{\cal K}_2 (z)$ is the
thermally averaged total decay rate.

These equations are the ones normally used in leptogenesis
calculations for the simplest case of one massive and one light
neutrino. However, compared with Eqs.~(\ref{eq:R}-\ref{eq:L2})
they involve the approximation of assuming Maxwell-Boltzmann
statistics and kinetic equilibrium. Particularly for the case
where $\Gamma/H(T=m_R) \sim 1$ this approximation is not
necessarily good.

When decays and inverse decays are included (as well as gauge
interactions of $L$ and $H$ which establish kinetic equilibrium for
$L,\bar{L}$, and $H$ individually, but which conserve $L-\bar{L}$)
there are essentially only two interesting parameters. All terms
that contribute to the development of densities are proportional to
either $H$ or $\Gamma$. This competition can be parameterised by the
single parameter $K=\Gamma/H(T=m_R)$, the decay parameter. The other
important parameter is the net decay asymmetry $\epsilon$.

In the present study we assume that the heavy neutrinos start with
zero abundance, i.e.\ that they are not equilibrated at high
temperatures through other interactions. We refer the reader to
\cite{Giudice:2003jh} for a thorough discussion of the various
possibilities for initial $R$ abundance.

In Figs. 1-3 we show a calculation of $n_R$ and the net asymmetry
for different values of $K$. At high temperatures the
equilibration rate of $R$ is much higher when the full Boltzmann
equation is used. The main reason for this can be seen in Fig.~4.
When $z < 1$ the low momentum states of $R$ are populated very
efficiently, while there is little population of high momentum
modes. This in turn means that the inverse decay $HL \to R$ is
suppressed relative to the decay process $R \to HL$.

To quantify this, we want to define a momentum dependant inverse
decay rate,
\begin{equation}
\Gamma(p_R)\equiv \frac{1}{f_{R,eq}(p_R)}\frac{d f_R}{d
t}|_{p=p_R}
\end{equation}
The efficiency with which $R$ states with momentum $p_R$ are
produced from the background (ignoring decays and Pauli Blocking)
is then given by
\begin{eqnarray}
\frac{\Gamma(p_R)}{H} & = & \frac{1}{f_{R,eq} H}\frac{d f_R}{d t}
= \frac{m_R \Gamma}{f_{R,eq} H E_R p_R}
\int_{(E_R-p_R)/2}^{(E_R+p_R)/2} dp_H f_H f_L. \\
&=& \frac{m_R T \Gamma}{H p_R E_R} \log\left[\sinh\left((E_R +
p_R)/2T\right)/\sinh \left((E_R - p_R)/2T\right)\right]\nonumber.
\end{eqnarray}
This function is sharply peaked at low momentum with the following
limiting behaviour
\begin{equation}
\frac{\Gamma(p_R)}{H} \to \cases{\frac{\Gamma}{H} \coth(m_R/2T)
\propto z^2 \coth(z/2) & $p_R \to 0$ \cr \frac{\Gamma m_R}{H
p_R^2} \left(p_R + T \log\left(\frac{2 p_R T}{m_R^2}\right)\right)
& $p_R \to \infty$}. \label{eq:limiting}
\end{equation}
This is a steeply increasing function of $z$ and at low
temperatures when $T << m_R$ the distribution of $R$ is in
complete equilibrium and $f_R << 1$ so that the Boltzmann
approximation is valid. Therefore the curves for $n_R/n_{\rm eq}$
approach each other in Figs.~1-3. Note, however, that the highest
$p_R$ modes are always out of equilibrium because of the
asymptotic $1/p_R$ term. In fact for given values of $z$ and $K$
there will be a limiting value of $p_R$ above which the
distribution will be out of equilibrium. In Fig.~5 we show the
value of $p_R/T$ for which $\Gamma(p_R)/H=1$ as a function of $z$,
for the specific case of $K=1$. At high temperatures no momentum
states are in equilibrium, whereas at low temperatures only
progressively higher momentum states are out of equilibrium. This
figure corresponds well to what is shown in Fig.~4. For $z=0.2$ no
states are in equilibrium, for $z=1$ states above $p_R/T \sim 1$
are not yet in equilibrium, and for $z=5$ almost all states have
been populated to their equilibrium value.

As can be seen in Figs.~1-3 the net asymmetry also grows much more
rapidly initially when the full Boltzmann equation is used, and
the change of sign occurs at higher temperature. However, at low
temperatures the differences are smaller.

\begin{figure}
\begin{center}
\includegraphics[width=0.6 \textwidth]{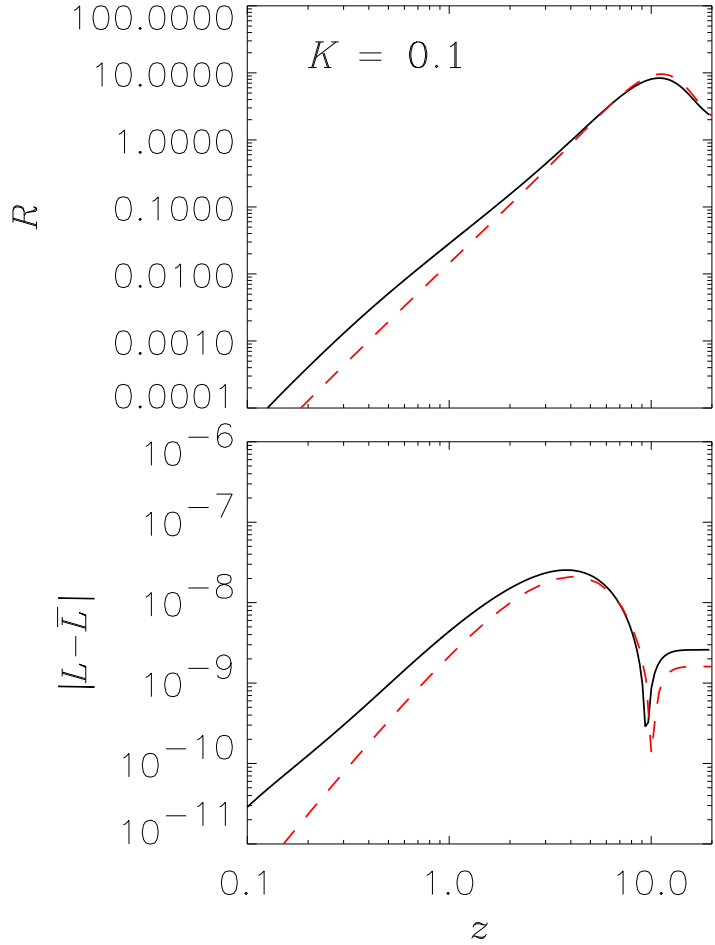}
\caption{Upper panel: The evolution of $n_R/n_{\rm eq}$ for the
full (solid line) and integrated (dotted line) Boltzmann
equations. Bottom panel: The evolution of $|n_{L-\bar{L}}|/T^3$
for the same cases. The calculation is for $K=0.1$ and
$\epsilon=10^{-6}$. }\label{fig:fig1}
\end{center}
\end{figure}

\begin{figure}
\begin{center}
\includegraphics[width=0.6 \textwidth]{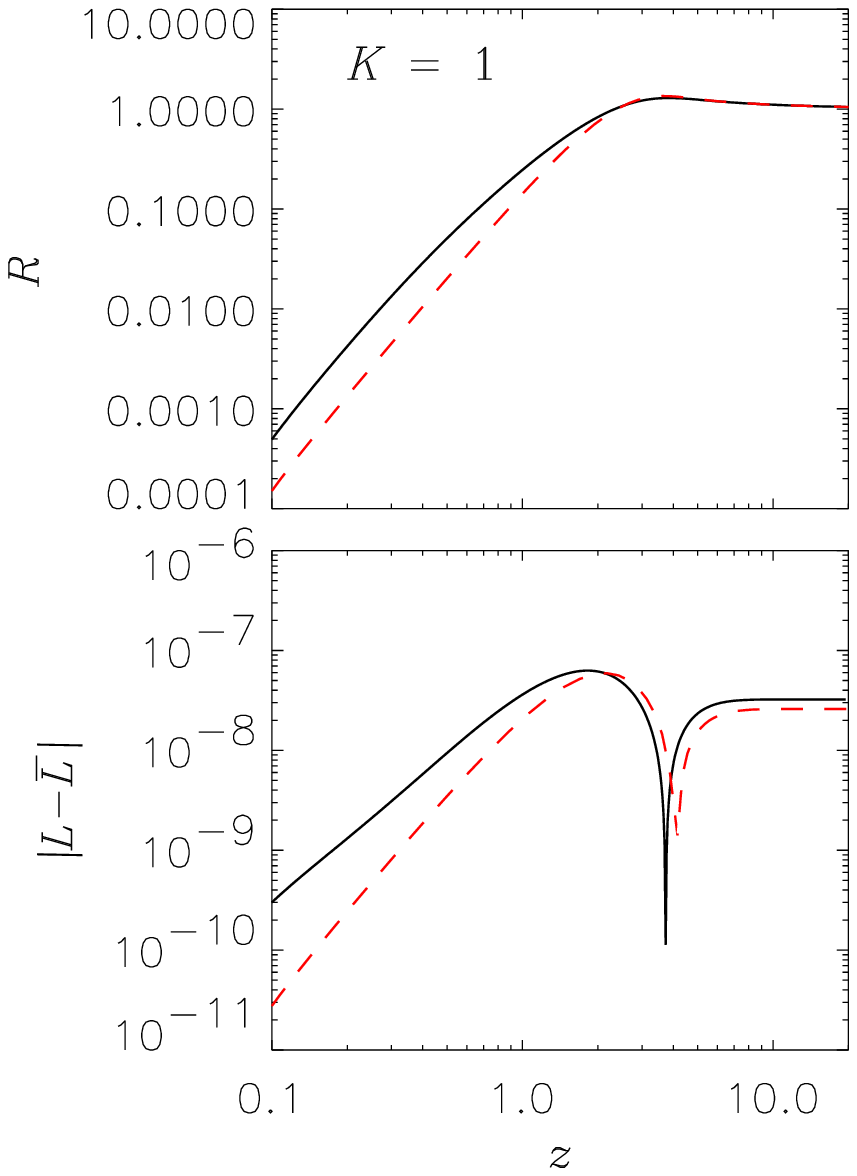}
\caption{Upper panel: The evolution of $n_R/n_{\rm eq}$ for the
full (solid line) and integrated (dotted line) Boltzmann
equations. Bottom panel: The evolution of $|n_{L-\bar{L}}|/T^3$
for the same cases. The calculation is for $K=1$ and
$\epsilon=10^{-6}$. }\label{fig:fig2}
\end{center}
\end{figure}

\begin{figure}
\begin{center}
\includegraphics[width=0.6 \textwidth]{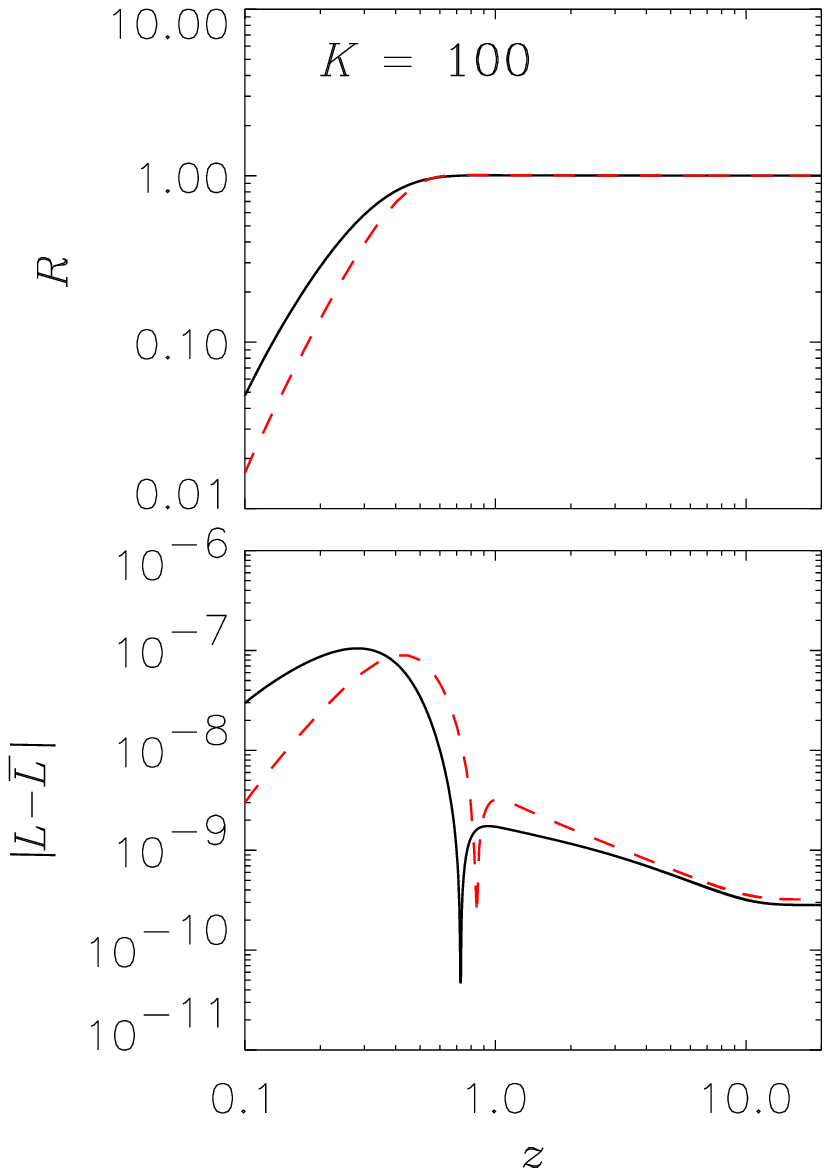}
\caption{Upper panel: The evolution of $n_R/n_{\rm eq}$ for the
full (solid line) and integrated (dotted line) Boltzmann
equations. Bottom panel: The evolution of $|n_{L-\bar{L}}/T^3|$
for the same cases. The calculation is for $K=100$ and
$\epsilon=10^{-6}$. }\label{fig:fig3}
\end{center}
\end{figure}

\begin{figure}
\begin{center}
\includegraphics[width=0.6 \textwidth]{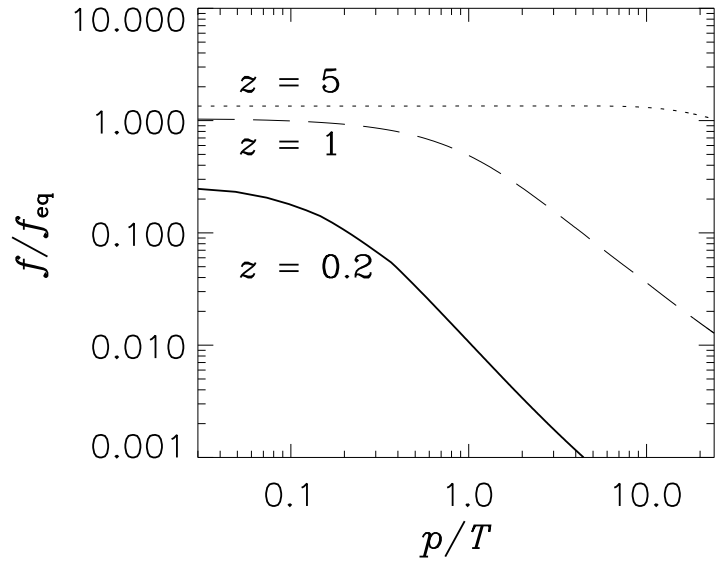}
\caption{The distribution function of $R$ relative to a chemical
equilibrium Fermi-Dirac distribution with the same temperature for
different values of $z$. The calculation is for
$K=1$.}\label{fig:fig4}
\end{center}
\end{figure}

\begin{figure}
\begin{center}
\includegraphics[width=0.6 \textwidth]{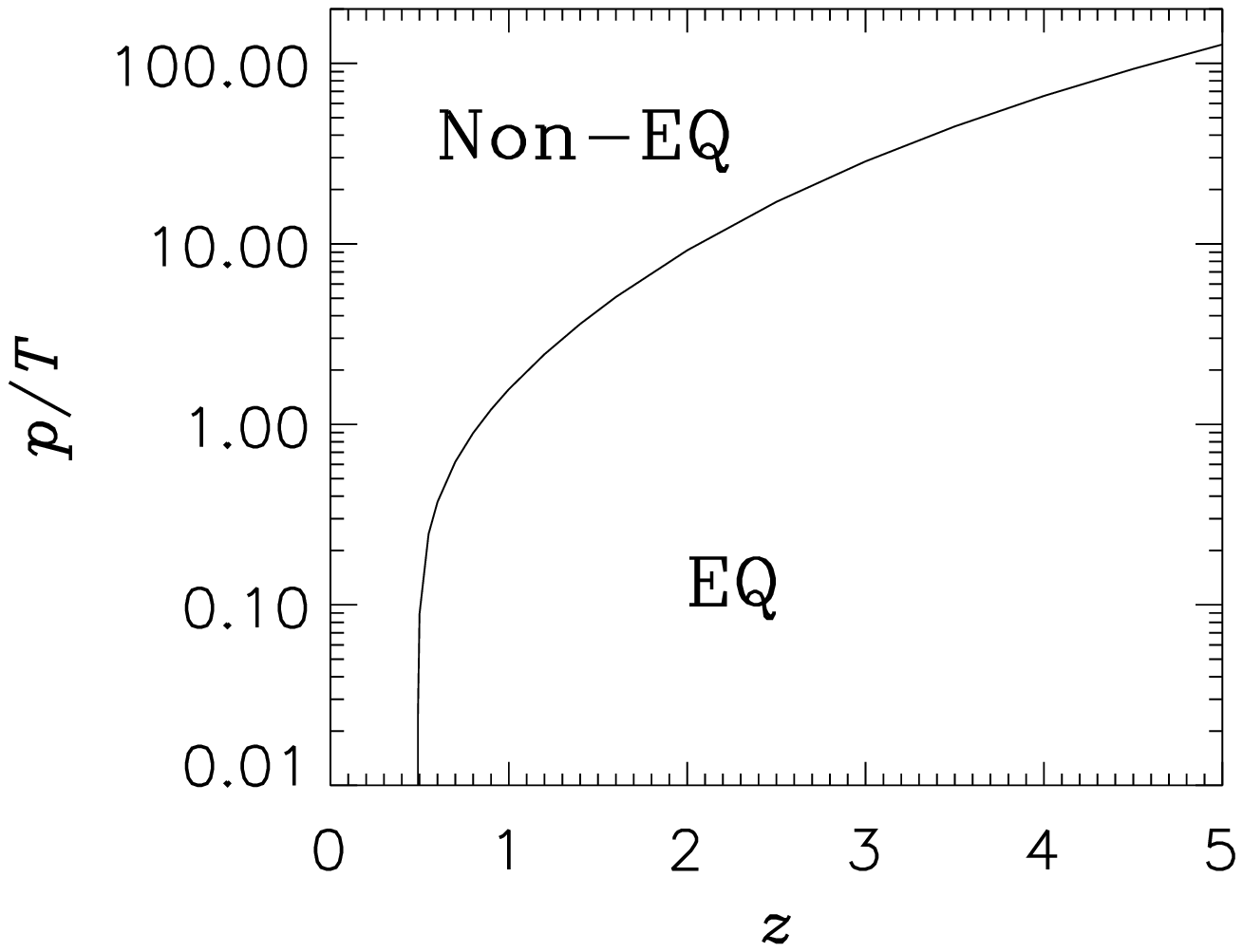}
\caption{The dividing line between equilibrium and non-equilibrium
for $K=1$. All momentum states above the line are out of
equilibrium.}\label{fig:fig5}
\end{center}
\end{figure}

In Fig. 6 we show the ratio of the final asymmetry in the two
cases
\begin{equation}
\frac{\eta_{\rm full}}{\eta_{\rm int}} =
\frac{\left.(n_L-n_{\bar{L}})/n_\gamma\right|_{\rm
full}}{\left.(n_L-n_{\bar{L}})/n_\gamma\right|_{\rm int}}
\end{equation}

As can be seen from Fig.~6 there can be a significant difference
in the asymmetry between the two approaches. For very high values
of $K$ the difference is small because the distribution is kept
very close to kinetic equilibrium (for a Fermi-Dirac distribution)
throughout the decay. Note, however, that the difference never
goes to zero. The reason is that the thermal decay width is
changed relative to the Maxwell-Boltzmann approximation when Pauli
blocking and stimulated emission factors are included (as also
noted by \cite{Giudice:2003jh}). For smaller values of $K$ the
difference increases and can be as large as 50\%. However, it
should be noted that our framework will break down for small
values of $K$ because we have not included scattering.

Another important point to note is that $\eta_{\rm full}/\eta_{\rm
int}$ does not depend on $\epsilon$ because the deviation from
kinetic equilibrium is governed completely by the total decay rate
(i.e.\ by $K$), not by the asymmetry, as long as $\epsilon << 1$.

We are only interested in the final value of the asymmetry since,
in order to maximise the baryon asymmetry for any given value of
$\epsilon$, we want leptogenesis to end before the conversion
through sphalerons end.

\begin{figure}
\begin{center}
\includegraphics[width=0.6 \textwidth]{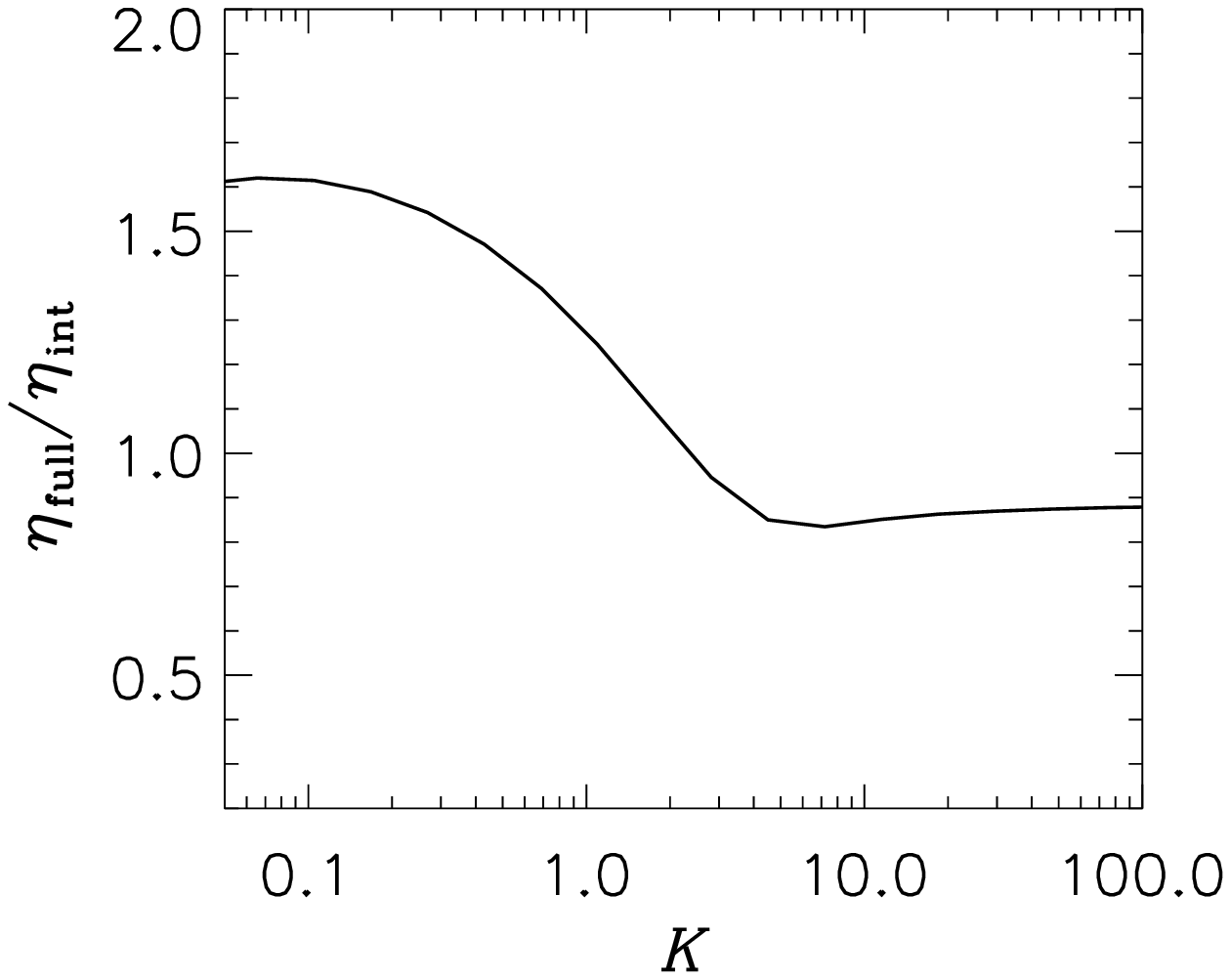}
\caption{The difference between the asymmetry in the full case and
the integrated case as a function of $K$.}\label{fig:fig6}
\end{center}
\end{figure}

%%%%%%%%%%%%%%%%%%%%%%%%%%%%%%%%%%%%%%%%%%%%%%%%%%%%%%%%%%%%%%%%%%%%%%
\section{Discussion} %%%%%%%%%%%%%%%%%%%%%%%%%%%%%%%%%%%%%%%%%%%%%%%%%
%%%%%%%%%%%%%%%%%%%%%%%%%%%%%%%%%%%%%%%%%%%%%%%%%%%%%%%%%%%%%%%%%%%%%%

We have solved the full Boltzmann equation for decays and inverse
decays to follow the generation of lepton asymmetry during
leptogenesis. When decays are semi-relativistic the difference
between using the full Boltzmann equation and the standard
integrated Boltzmann equation can be very large. However, at low
temperature where washout dominates the difference is relatively
modest.

The difference at $K \gtrsim 1$ is less than about 30\% between
the two approaches. For smaller $K$ the difference can be larger.
However, this is the regime where $1 \leftrightarrow 2$ processes
do not dominate, i.e.\ outside the regime in which our framework
is valid.

The conclusion is that the Boltzmann approximation yields results
in the strong washout regime which are accurate to at least 30\%.
For $K > 5$ the difference is not larger than 15\%.

%%%%%%%%%%%%%%%%%%%%%%%%%%%%%%%%%%%%%%%%%%%%%%%%%%%%%%%%%%%%%%%%%%%%%%
\section*{Acknowledgments} %%%%%%%%%%%%%%%%%%%%%%%%%%%%%%%%%%%%%%%%%%%
%%%%%%%%%%%%%%%%%%%%%%%%%%%%%%%%%%%%%%%%%%%%%%%%%%%%%%%%%%%%%%%%%%%%%%

SH acknowledges support from the Alexander von Humboldt Foundation
through a Friedrich Wilhelm Bessel Award. We thank Pasquale Di
Bari and Alessandro Strumia for valuable comments.

\vspace*{2cm}

%%%%%%%%%%%%%%%%%%%%%%%%%%%%%%%%%%%%%%%%%%%%%%%%%%%%%%%%%%%%%%%%%%%%%%
\section*{References} %%%%%%%%%%%%%%%%%%%%%%%%%%%%%%%%%%%%%%%%%%%%%%%%
%%%%%%%%%%%%%%%%%%%%%%%%%%%%%%%%%%%%%%%%%%%%%%%%%%%%%%%%%%%%%%%%%%%%%%

\end{document}